\documentclass[journal]{IEEEtran}
\usepackage[utf8]{inputenc}
\usepackage{graphicx}
\usepackage{caption}
\usepackage{subcaption}
\usepackage{amsmath,amssymb,amsfonts}
\usepackage[hyperfootnotes=false]{hyperref}
\usepackage{array}
\usepackage{float}
\usepackage{adjustbox}
\usepackage{booktabs}
\usepackage[acronym]{glossaries}

\usepackage[backend=biber, style=ieee, sorting=none, dashed=false, doi=false]{biblatex}

\hypersetup{
    colorlinks,
    linkcolor=[rgb]{0.1,  0.1,  0.1},  %blue={0.1,  0.4,  0.7},
    citecolor=[rgb]{0.0,  0.75, 0.0},
    urlcolor= [rgb]{0.6,  0.0,  0.6}
}

\begin{document}

%%%%%%%%%%%%%%%%%%%%%%%%%%%%%%%%%%%%%%%%%%%%%%%%%%%%%%%%%%%%%%%%%%%%%%%%%%%%%%%%%%%%%%%%%

\title{Complete Inertial Pose Dataset: from raw measurements to pose with low-cost and high-end MARG sensors}
%\title{Complete Inertial Pose Dataset: wearable pipeline from raw measurements to pose with low-cost and high-end MARG sensors}

\author{
    \IEEEauthorblockN{
        Manuel~Palermo*\thanks{* Corresponding Author (a76886@alunos.uminho.pt).}\IEEEauthorrefmark{2}\IEEEauthorrefmark{4},
        Sara~Cerqueira\IEEEauthorrefmark{2}\IEEEauthorrefmark{4},
        João~André\IEEEauthorrefmark{2}\IEEEauthorrefmark{4},
        António~Pereira\IEEEauthorrefmark{2}\IEEEauthorrefmark{4},
        Cristina~P.~Santos\IEEEauthorrefmark{2}\IEEEauthorrefmark{4}
    }\\
    \thanks{\IEEEauthorrefmark{2}{Center for Microelectromechanical Systems (CMEMS), University of Minho, Guimarães, Portugal}}
    \thanks{\IEEEauthorrefmark{4}{LABBELS – Associate Laboratory, Guimarães, Portugal}}
    %
    %\thanks{E-mail address: a76886@alunos.uminho.pt (Manuel Palermo) sara.cerqueira@mail.pt (Sara Cerqueira), joao.andre@mail.pt (João André), antonio.pereira@mail.pt (António Pereira), Cristina P. Santos (cristina@dei.uminho.pt)}
}

\maketitle

%%%%%%%%%%%%%%%%%%%%%%%%%%%%%%%%%%%%%%%%%%%%%%%%%%%%%%%%%%%%%%%%%%%%%%%%%%%%%%%%%%%%%%%%%

\begin{abstract}
    % motivation
    The use of wearable technology for posture monitoring has been expanding due to its low-intrusiveness and compliance with daily use requirements. However, there are still open challenges limiting its widespread use, especially when dealing with low-cost systems. Most solutions falls either into fully functioning commercial products with high costs, or ad-hoc solutions with lower performance. Moreover, there are few datasets available, from which complete and general solutions can be derived.
    % introduction
    This work presents 2 datasets, containing low-cost and high-end Magnetic, Angular Rate, and Gravity (MARG) sensor data respectively. It provides data for the analysis of the complete inertial pose pipeline, from raw measurements, to sensor-to-segment calibration, multi-sensor fusion, skeleton kinematics, to the complete human pose.
    % methods
    Multiple trials were collected with 21 and 10 subjects respectively, performing 6 types of sequences (ranging from calibration, to daily-activities and random movements). It presents a high degree of variability and complex dynamics with almost complete range-of-motion, while containing common sources of error found on real conditions.
    % results
    This amounts to 3.5M samples, synchronized with a ground-truth inertial motion capture system at 60hz. A simple end-to-end inertial pose method was briefly described and used to validate the quality of the data in both acquisitions.
    %conclusion
    This database may contribute to assess, benchmark and develop novel algorithms for each of the pipelines’ processing steps, with applications in classic or data-driven inertial pose estimation algorithms, human movement understanding and forecasting and ergonomic assessment in industrial or rehabilitation settings.
    % dataset availability
    All the data is freely available on an online database and accompanied with code to process and analyze the complete data pipeline.
\end{abstract}


% \begin{highlights}
% \item 
% \item 
% \item
% \end{highlights}

% Note that keywords are not normally used for peer-review papers.
\begin{IEEEkeywords}
    dataset, human pose estimation, IMU, MARG, sensor fusion
\end{IEEEkeywords}

\IEEEpeerreviewmaketitle

%%%%%%%%%%%%%%%%%%%%%%%%%%%%%%%%%%%%%%%%%%%%%%%%%%%%%%%%%%%%%%%%%%%%%%%%%%%%%%%%%%%%%%%%%

%\linenumbers % start numbering pages (cannot be called sooner or it breaks author footnotes)

% add acronyms (not all were used)
\newacronym{COM}{COM}{Center-of-Mass}
\newacronym{HRI}{HRI}{Human-Robot Interaction}
\newacronym{MoCap}{MoCap}{Motion Capture}
\newacronym{IMU}{IMU}{Inertial Measurement Unit}
\newacronym{MARG}{MARG}{Magnetic, Angular Rate, and Gravity}
\newacronym{NWU}{NWU}{North-West-Up}
\newacronym{FPS}{FPS}{Frames per Second}
\newacronym{GT}{GT}{Ground-Truth}
\newacronym{FK}{FK}{Forward Kinematics}
\newacronym{EKF}{EKF}{Extended Kalman Filter}
\newacronym{DL}{DL}{Deep learning}
\newacronym{NN}{NN}{Neural Network}
\newacronym{RNN}{RNN}{Recurrent Neural Network}
\newacronym{CNN}{CNN}{Convolutional Neural Network}
\newacronym{GNN}{GNN}{Graph Neural Network}
\newacronym{MAE}{MAE}{Mean Absolute Error}
\newacronym{MSE}{MSE}{Mean Squared Error}
\newacronym{QAD}{QAD}{Quaternion Angle Distance}
\newacronym{HPE}{HPE}{Human Pose Estimation}
\newacronym{ROS}{ROS}{Robot Operating System}
\newacronym{ONNX}{ONNX}{Open Neural Network Exchange}
\newacronym{API}{API}{Application Programming Interface}
\newacronym{SDK}{SDK}{Software development kit}
\newacronym{SotA}{SotA}{State-of-the-Art}

\section{Introduction}
\label{sec:intro}

% Context
Inertial-based wearable technology is being quickly adopted for many applications requiring estimation of body configuration in real-time (e.g. human motion analysis, ergonomic assessment, virtual interaction, etc...), given their relative low-cost and ease of use, while presenting reasonable accuracy. Moreover, the light-source and field-of-view independence allows the user to be unconstrained from predefined locations, contributing to more natural movements, while avoiding data privacy issues, contrary to camera-based solutions \cite{camomilla2018imutrends, lopez2016imusreview}.

% Motivation
Despite their widespread adoption, multiple challenges still remain, mainly when it comes to applications requiring high accuracy (e.g. rehabilitation assessment) and dealing with whole-body dynamics with joints that have multiple degrees of freedom \cite{lopez2016imusreview} in real world settings, given the:
\textbf{i)} complex dynamical movements over the whole range of joint motion;
\textbf{ii)} offset errors when converting from the sensor reference frame to the body segments frame;
\textbf{iii)} possibility of magnetic interference (when using magnetometer fusion) or yaw drift (when not using magnetometer fusion).

% moved here so it appears on desired place on paper
\begin{figure}[t]
    \centering
    \begin{subfigure}[b]{0.26\textwidth}
        \centering
        \includegraphics[height=4cm]{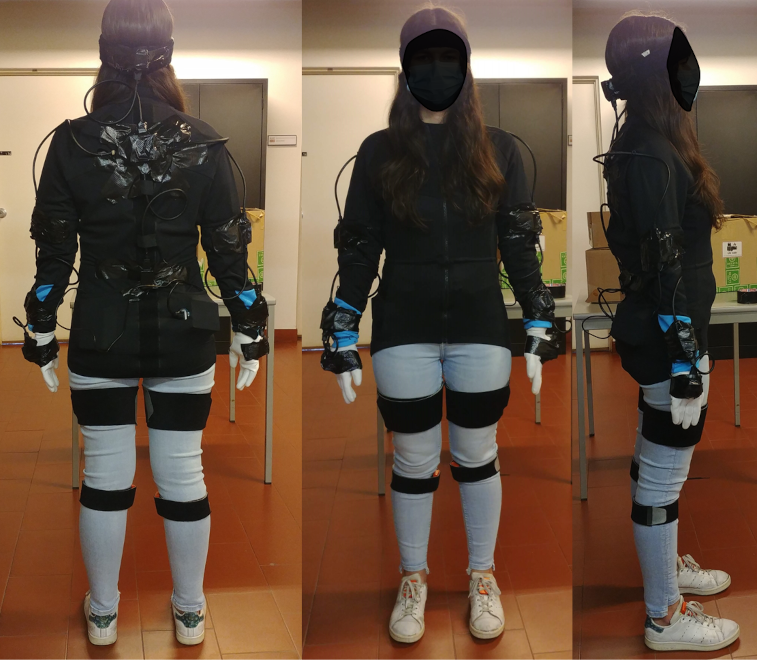}
        \caption{Ergowear}
        \label{fig:acquisition_setup_ergowear}
    \end{subfigure}
    \begin{subfigure}[b]{0.22\textwidth}
        \centering
        \includegraphics[height=4cm]{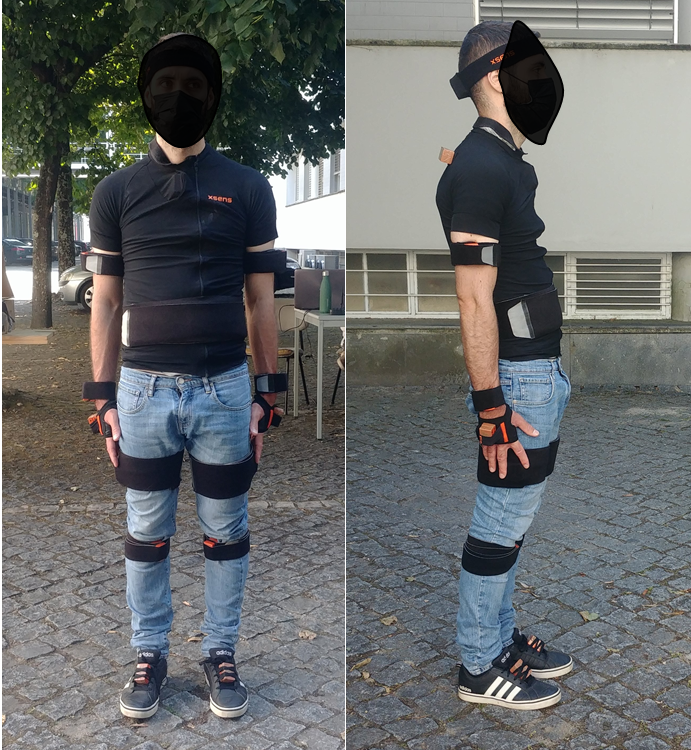}
        \caption{MTwAwinda}
        \label{fig:acquisition_setup_mtmanager}
    \end{subfigure}
    \caption{\textbf{(a)} Participant instrumented with Xsens MTw Awinda and Ergowear during experimental data collection D1. \textbf{(b)} Participant instrumented with two Xsens MTw Awinda systems on top of each other, during experimental data collection D2.}
    \label{fig:acquisition_setup}
\end{figure}

% Related work
Most research has been focused on just one of these aspects, an it is missing unified datasets/benchmark with ample amounts of data, needed to develop and evaluate complete products \cite{camomilla2018imutrends, lopez2016imusreview}.
Some datasets for inertial pose estimation analysis exist in the literature. However, these present some limitations which prevent them from being directly applicable to real-world situations:
\textbf{i)} most datasets \cite{Huang2018dipimu, Trumble2017totalcapture, Luo2020gaitdatabase} use the commercial Xsens Inertial \gls{MoCap} system's (Xsens Technologies, B.V., The Netherlands \cite{roetenberg2009xsens}) hardware/software for data acquisition. Therefore, data has already been pre-processed to a higher quality level. However, this data is not representative of the lower grade hardware/software, used in most real-world applications \cite{camomilla2018imutrends};
\textbf{ii)} These datasets are geared towards developing novel inertial fusion algorithms. Notwithstanding, they ignore the rest of the pipeline involved to create working pose estimation solutions from inertial data (e.g. sensor-to-segment calibration, magnetic-rejection, etc...), which are necessary when considering end products and which are still open investigation problems \cite{Choe2019s2s, lopez2016imusreview, camomilla2018imutrends};
\textbf{iii)} most papers looking into inertial motion estimation have focused on using a low amount of sensors, attached to individual limbs, mostly ignoring multi-sensor fusion. Hence, they present relatively simple whole body dynamics, with subjects focusing on lower- or upper-body movements, with little full range-of-motion \cite{Luo2020gaitdatabase, lopez2016imusreview, camomilla2018imutrends};
\textbf{iv)} Given the lack of large scale full-body inertial pose datasets, \textit{Huang et al.} \cite{Huang2018dipimu} have relied on virtual \gls{IMU} data obtained from large scale \gls{MoCap} databases \cite{Mahmood2019Amass}, by derivation of linear and angular segment positions to respectively obtain acceleration and gyroscope data, similarly to an \gls{IMU}. Nevertheless, this data does not contain real sensor characteristics (ex. noise, bias, etc...), and no magnetometer data is available, which is necessary in real systems to prevent yaw drift in long time usage \cite{lopez2016imusreview}.

% Selling points
In order to address some of these limitations, we present a novel dataset for inertial pose estimation. It differs from existing literature by providing: 
\textbf{\textit{(i)}} Access to the full data pipeline, from raw data, to sensor-to-segment calibration, multi-sensor fusion, skeleton-kinematics, to the complete human pose;
\textbf{\textit{(ii)}} High variability and fast dynamic trials with almost complete range-of-motion;
\textbf{\textit{(iii)}} Trials focusing on upper-body daily activities with natural movements;
\textbf{\textit{(iv)}} Two datasets with 2 types of \gls{MARG} sensors, one high-end (Xsens MTw) and another low-cost (MPU9250, collected with a similar acquisition protocol to enable benchmarking;
\textbf{\textit{(v)}} Large amount of data ($\sim$3.5M samples at 60Hz) synchronized with a \gls{GT} Inertial \gls{MoCap} system.

This article is divided into 4 sections: 
\textbf{\textit{i)}} the methodology used to acquire and process the data;
\textbf{\textit{ii)}} the database created with the data and its structure; 
\textbf{\textit{iii)}} brief validating results with baseline values; 
\textbf{\textit{iv)} }indications on where to use and access all the data and code used in this work.

%%%%%%%%%%%%%%%%%%%%%%%%%%%%%%%%%%%%%%%%%%%%%%%%%%%%%%%%%%%%%%%%%%%%%%%%%%%%%%%%%%%%
%%%%%%%%%%%%%%%%%%%%%%%%%%%%%%%%%%%%%%%%%%%%%%%%%%%%%%%%%%%%%%%%%%%%%%%%%%%%%%%%%%%%
\section{Methods}
\label{sec:methods}

This section lays out the process or data collection of the 2 datasets: Ergowear dataset (D1) and Mtw Awinda dataset (D2).
The acquisition of the first dataset (D1), concerning low cost sensor data, was focused on general upper-body indoor usage, which might be found in rehabilitation, office or industrial settings. This wearable \cite{resende2021ergowear} provides representative data of low-cost systems and \textit{ad-hoc} software, which can easily be built for many applications.
Nevertheless, the use of such systems poses some challenges in terms of achieving high accuracy. This results not only from the low cost hardware (relatively high noise and bias instability), but also from the lack of supporting technologies that exist on most higher-end sensors, namely the use of Strapdown Integration \cite{Paulich2018mtw, xsens_sdi}, self-calibrating sensors, magnetic rejection, anti-slide straps, among others. For this reason, we conducted a second data acquisition (D2), following a similar protocol using the high-end Xsens MTw Awinda (Xsens Technologies, B.V., The Netherlands) \cite{Paulich2018mtw, roetenberg2009xsens} sensors instead, which use these technologies to produce high quality "raw" sensor data. These two datasets allow a more comprehensive test of new algorithms and to benchmark them on different conditions, i.e., near of ideal hardware (Xsens) and easily accessible hardware.

%%%%%%%%%%%%%%%%%%%%%%%%%%%%%%%%%%%%%%%%%%%%%%%%%%%%%%%%%%%%%%%%%%%%%%%%%%%%%%%%
\subsection{Participants}
\label{sec:methods_participants}

\begin{table}[t]
    \caption{Ergowear data acquisition (D1), participants' main anthropometric data.}
    \label{tab:participants_metadata_d1}
    %\maxsizebox{0.90\textwidth}{!}{
    \begin{tabular}{c c c c c}
        \toprule
        \textbf{Participant} & \textbf{\begin{tabular}[c]{@{}c@{}}Gender\\(M/F)\end{tabular}} & \textbf{\begin{tabular}[c]{@{}c@{}}Age\\(years)\end{tabular}} & \textbf{\begin{tabular}[c]{@{}c@{}}Body mass\\(kg)\end{tabular}} & \textbf{\begin{tabular}[c]{@{}c@{}}Body height\\(cm)\end{tabular}} \\ \midrule
        00 & M & 24 & 67 & 180 \\
        01 & M & 22 & 64 & 174 \\
        02 & F & 24 & 50 & 151 \\
        03 & F & 28 & 68 & 159 \\
        04 & F & 27 & 52 & 157 \\
        05 & M & 30 & 70 & 174 \\
        06 & F & 28 & 53 & 162 \\
        07 & M & 24 & 75 & 170 \\
        08 & M & 22 & 70 & 176 \\
        09 & M & 26 & 80 & 177 \\
        10 & M & 23 & 72 & 176 \\
        11 & F & 22 & 59 & 160 \\
        12 & M & 24 & 65 & 171 \\
        13 & M & 25 & 65 & 175 \\ 
        14 & M & 26 & 70 & 181 \\
        15 & M & 25 & 64 & 175 \\
        16 & M & 27 & 74 & 181 \\
        17 & M & 22 & 64 & 176 \\
        18 & M & 22 & 73 & 168 \\
        19 & M & 26 & 71 & 181 \\
        20 & F & 24 & 61 & 171 \\ \midrule
           & \textbf{15M, 6F} & \textbf{25.0 ($\pm2.3$)} & \textbf{66.0 ($\pm7.8$)} & \textbf{171 ($\pm8.6$)} \\ \bottomrule
    \end{tabular}
    %}
\end{table}

\begin{table}[t]
    \caption{MTwAwinda data acquisition (D2), participants' main anthropometric data. Some of the participants were also present in the first data collection.}
    \label{tab:participants_metadata_d2}
    \maxsizebox{0.90\textwidth}{!}{
    \begin{tabular}{c c c c c}
        \toprule
        \textbf{Participant} & \textbf{\begin{tabular}[c]{@{}c@{}}Gender\\(M/F)\end{tabular}} & \textbf{\begin{tabular}[c]{@{}c@{}}Age\\(years)\end{tabular}} & \textbf{\begin{tabular}[c]{@{}c@{}}Body mass\\(kg)\end{tabular}} & \textbf{\begin{tabular}[c]{@{}c@{}}Body height\\(cm)\end{tabular}} \\ \midrule
        00 & M & 24 & 76 & 170 \\
        01 & F & 24 & 61 & 171 \\
        02 & M & 24 & 65 & 171 \\
        03 & M & 22 & 70 & 176 \\
        04 & M & 24 & 67 & 180 \\
        05 & M & 30 & 70 & 174 \\
        06 & F & 23 & 63 & 168 \\
        07 & M & 26 & 71 & 181 \\
        08 & F & 21 & 51 & 165 \\
        09 & F & 24 & 50 & 153 \\ \midrule
           & \textbf{6M, 4F} & \textbf{24.2 ($\pm2.6$)} & \textbf{64.4 ($\pm8.5$)} & \textbf{171 ($\pm8.1$)} \\ \bottomrule
    \end{tabular}
    }
\end{table}

Healthy participants from the University of Minho academic community were contacted to participate in the 2 studies. Both times, they were provided with the study’s goal and details, protocol, and duration. To select and recruit the participants, the following inclusion criteria was defined: \textbf{i)} have more than 18 years; \textbf{ii)} present full motion control and no clinical history of motor injuries.

21 healthy subjects (15 males and 6 females; body mass: 66.0 $\pm$ 7.8 kg; body height: 171 $\pm$ 8.6 cm; age: 25.0 $\pm$ 2.3 years-old) and 10 healthy subjects (6 males and 4 females; body mass: 64.4 $\pm$ 8.5 kg; body height: 171 $\pm$ 8.6 cm; age: 24.2 $\pm$ 2.6 years-old) were recruited for the first and second data collection, respectively.
All participants accepted to participate, voluntarily and provided their written and informed consent to participate in the study, according to the ethical conduct defined by the University of Minho Ethics Committee in Life and Health Sciences (CEICVS 006/2020), that follows the standard set by the declaration of Helsinki and the Oviedo Convention. Participants’ rights were preserved and, as such, personal information that could identify them remained confidential and it is not provided in this dataset.

\subsection{Instrumentation and Data Collection}
\label{sec:methods_instrumentation}

All data was collected in the School of Engineering of University of Minho.

For D1, the data collection was conducted inside a conference room, selected to to present natural and representative conditions for indoor environments, and consisted of a standard conference room with tables, cabinets, chairs, among others, and is expected to contain a non-homogeneous magnetic field.

The D2 data collection was conducted in an empty outdoors parking lot to minimize the presence of magnetic interference in the data. This second acquisition tries to follow a best case scenario, providing a representative estimate of what can be achieved given the right conditions.
In both protocols, the participants were instructed to wear sneakers and tight-fitting clothes, such as tight jeans or leggings and shirts or strap tops. 

For D1, the instrumentation of the participants occurred as follows: 
\textbf{i)} donning of the full-body inertial motion tracking system MTw Awinda, used as \gls{GT}. The seventeen \gls{MARG} sensors were placed following the manufacturer's guidelines \cite{xsens_placement_guidelines}, on the head, shoulders, chest, arms, forearms, wrist, waist, thighs, shanks, and feet, secured with straps; 
\textbf{ii)} donning of the Ergowear smart garment \cite{resende2021ergowear}. This system is still in prototype phase and is embedded with 9 low-cost \gls{MARG} sensors (MPU9250, Adafruit Industries), placed on the upper-body: hands, forearms, arms, head,  upper-back (Around T4 vertebrae) and lower-back. After donning the Ergowear, the responsible investigator placed duct tape over the sensors, to reduce the probability of sensor sliding during the experimental protocol. 
Figure \ref{fig:acquisition_setup_ergowear} illustrates a subject instrumented according to this experimental protocol.

For D2, each participant was instrumented in a similar way (Figure \ref{fig:acquisition_setup_mtmanager}), as follows:
\textbf{i)} donning of the full-body inertial motion tracking system MTw Awinda, used as \gls{GT}. The seventeen \glspl{MARG} sensors were placed following the manufacturer's guidelines \cite{xsens_placement_guidelines}, on the head, shoulders, chest, arms, forearms, wrist, waist, thighs, shanks, and feet, secured with straps; 
\textbf{ii)} donning of a second full-body inertial motion tracking system MTw Awinda, placing the seventeen sensors precisely on top of the first system's and secured with straps. The Xsens MtManager (Xsens Technologies, B.V., The Netherlands) acquisition software was used to acquire the raw sensor data in this case.

In both acquisitions, all sensors were always placed by the same researcher, ensuring repeatability in the instrumentation procedure and, thus, minimizing errors caused by misplacement.

% moved here so it appears on desired place on paper
\begin{figure*}[t]
    \centering
    \begin{subfigure}[b]{0.26\textwidth}
        \raggedright
        \includegraphics[height=5cm]{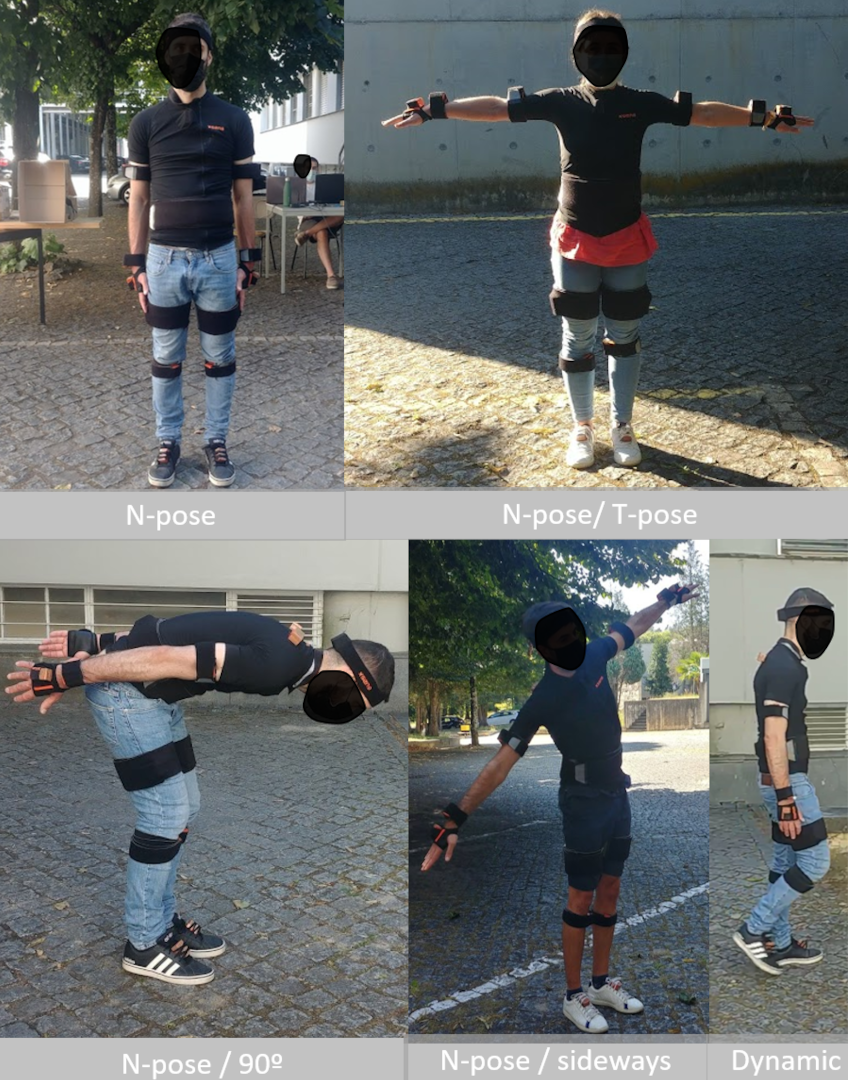}
        \caption{Calibration}
    \end{subfigure}
    \begin{subfigure}[b]{0.54\textwidth}
        \raggedleft
        \includegraphics[height=5cm]{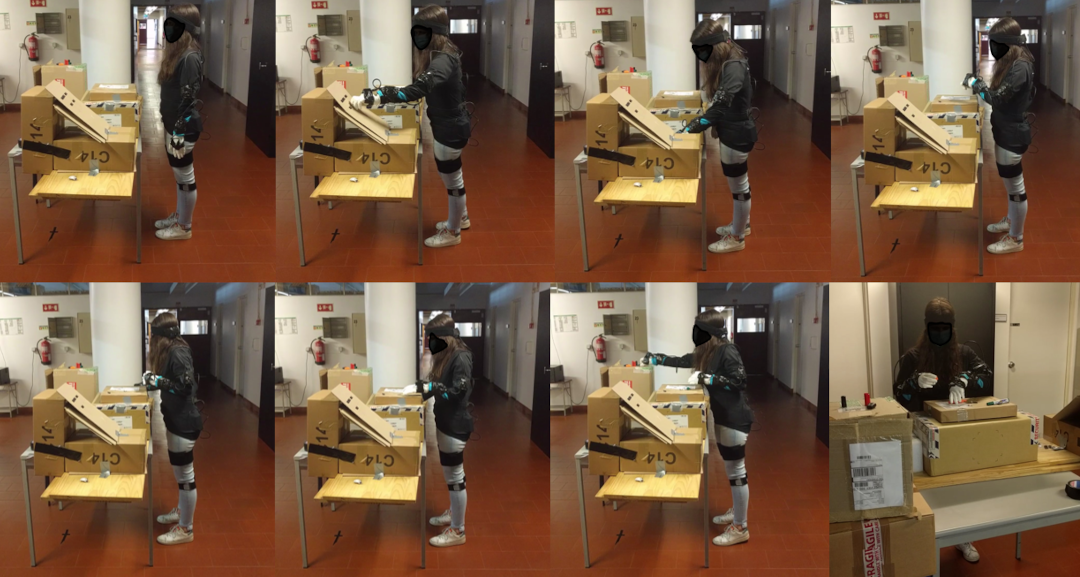}
        \caption{Task}
    \end{subfigure}
    %%%%%
    \begin{subfigure}[b]{0.38\textwidth}
        \raggedright
        \includegraphics[height=5cm]{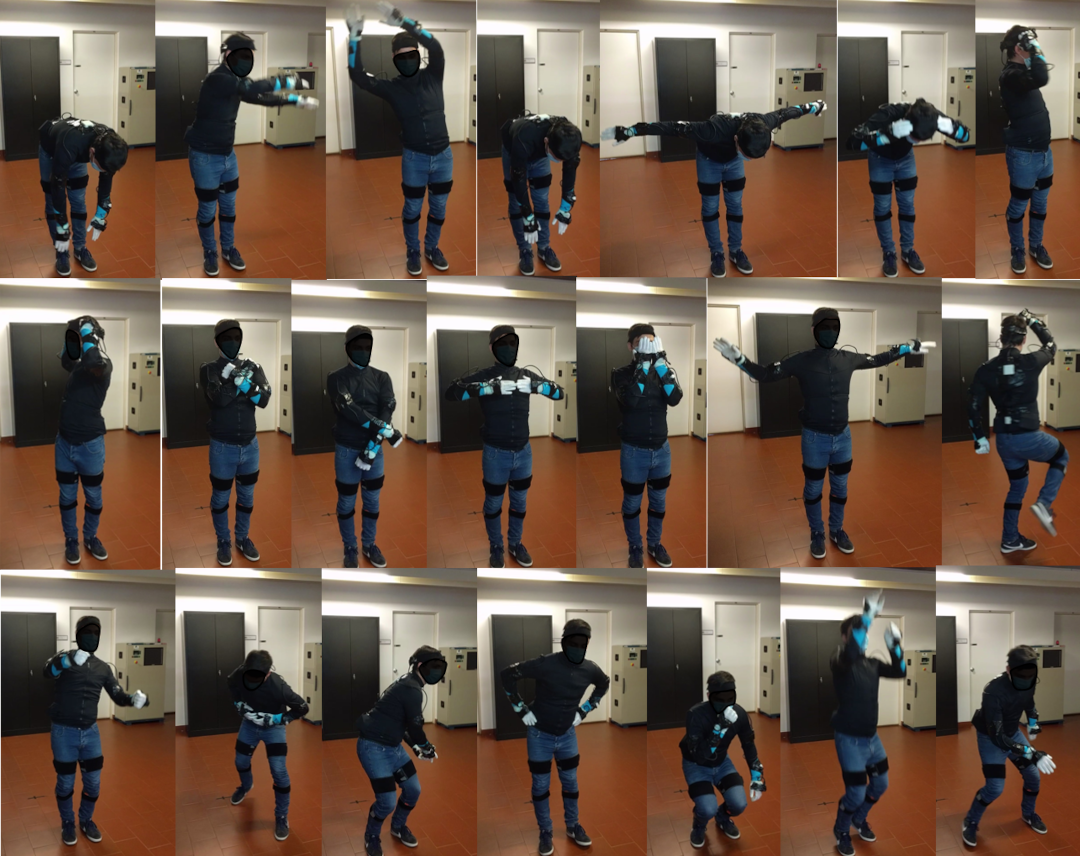}
        \caption{Sequence}
    \end{subfigure}
    \begin{subfigure}[b]{0.42\textwidth}
        \raggedleft
        \includegraphics[height=5cm]{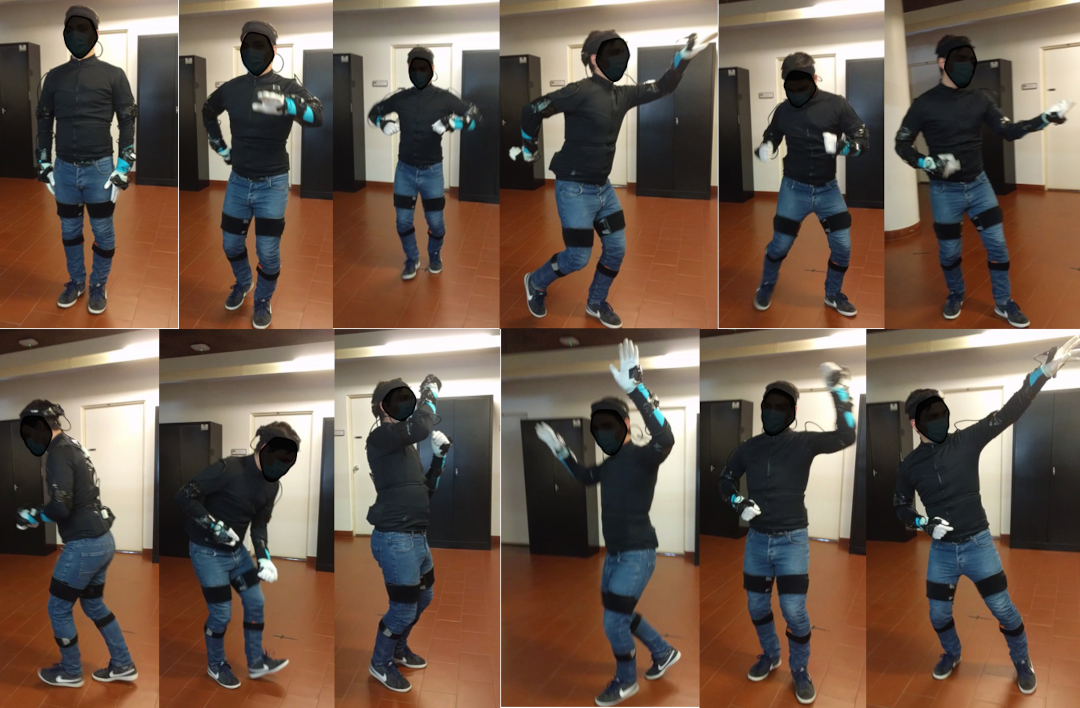}
        \caption{Random}
    \end{subfigure}
    %%%%%
    \begin{subfigure}[b]{0.50\textwidth}
        \raggedright
        \includegraphics[height=5cm]{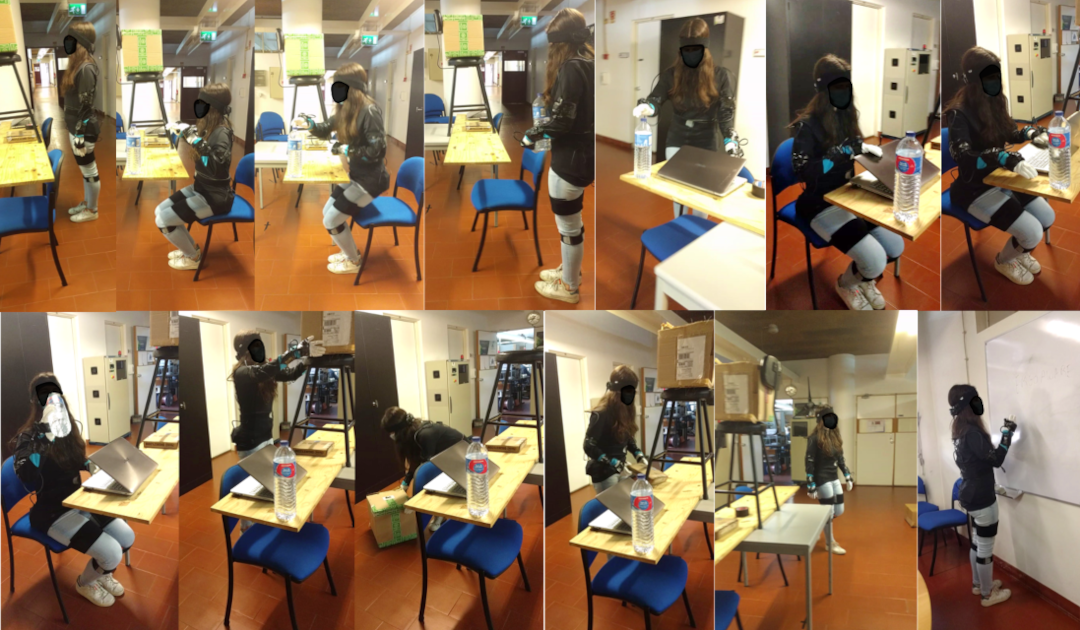}
        \caption{Circuit}
    \end{subfigure}
    \begin{subfigure}[b]{0.30\textwidth}
        \raggedleft
        \includegraphics[height=5cm]{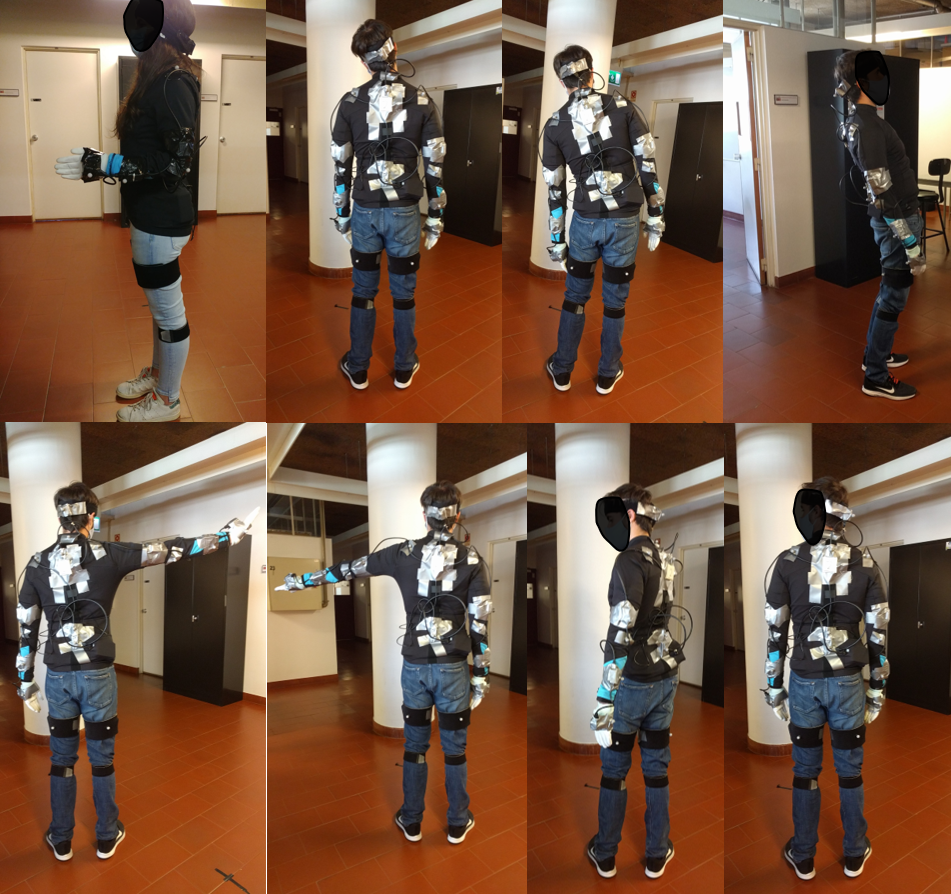}
        \caption{Validation}
    \end{subfigure}
    \caption{Frames, from video records captured during data acquisition, illustrative of datasets' movements for each of the 6 types of movements: Calibration (\textbf{a}), Task (\textbf{b}), Sequence (\textbf{c}), Random (\textbf{d}), Circuit (\textbf{e}) and Validation (\textbf{f}).}
    \label{fig:data_collection_sequences}
\end{figure*}

On both acquisitions, \gls{GT} data was synchronized and collected simultaneously, to evaluate any algorithmic solution developed. This was provided by the commercial \gls{MoCap} system (Xsens Awinda). This system uses the MTw Awinda sensors with the the Xsens Analyze software (Xsens Technologies, B.V., The Netherlands, validated in \cite{al2018inertial}), that integrates a biomechanical human model (MVN BIOMECH) to perform multi-sensor data fusion, along with better calibration routines, resulting in more accurate data.

Data collection included: \textbf{\textit{i)}} \gls{GT} kinematic data, namely sensors’ free acceleration, magnetic field, and orientation; segments’ orientation, position, velocity, and acceleration; and joints’ angles. These were acquired at 60 Hz using the MVN Analyze software and can be consulted on the user manual \cite{xsens_mvn_usermanual}; and \textbf{\textit{ii)}} \gls{MARG} raw sensor data obtained from the sensors in both cases. All data was synchronized in time using a hardware trigger.

%%%%%%%%%%%%%%%%%%%%%%%%%%%%%%%%%%%%%%%%%%%%%%%%%%%%%%%%%%%%%%%%%%%%%%%%%%%%%%%%
\subsection{Experimental protocol}
\label{sec:methods_protocol}

After sensorizing the subjects, their anthropometric data was collected according to the Xsens guidelines \cite{xsens_acquisition_guidelines}. These dimensions were introduced on the MVN Analyze to adjust the software’s biomechanical model (MVN BIOMECH) to the participant. Then, the MVN BIOMECH was calibrated in \emph{N-Pose}. The calibration’s quality was ensured for each subject, performing the necessary repetitions until a good quality was achieved.

The subjects were then instructed to perform 6 types of movement sequences (Figure \ref{fig:data_collection_sequences}), each focusing on different movement dynamics. Each type was repeated 3 and 5 times, respectively, for the D1 and D2 acquisitions. The repetitions only imply that the type of movements are the same, as people will move freely with very different dynamics in most cases, especially in the "random" trials. The 6 types of movement sequences are described below:

\begin{enumerate}
    \item \textbf{Calibration} - sensor to segment calibration sequences focused on the Upper-body. Includes static poses, functional movements, and the dynamic calibration similar to the one used in Xsens calibration (moving forward and then returning). D1 dataset only contains the dynamic calibration.
    \item \textbf{Task} – movements found on an simplified factory assembly task, where the user grabs small parts to assemble a product, using both hands. The movement contains reaching movements with both arms, with relatively small variability, and with the user standing roughly in the same position.
    \item \textbf{Circuit} – movements which mimic office dynamics, while interacting with some objects. The user sits in an office chair, works on the computer, moves some boxes around, walks and then writes something on a board. These trials might contain additional magnetic  interference from interacted objects.
    \item \textbf{Sequence} – sequence of movements which try to cover the full range-of-motion of human upper-body joints.
    \item \textbf{Random} – user performs free movements for $\sim$1.5 minutes. Example of movements produced include dancing, imitating sports, randomly moving all limbs, pretending to do some task, running, etc…
    \item \textbf{Validation} – isolated movements containing maximum range of motion for each upper-body joint individually. Only collected on the Ergowear acquisition.
\end{enumerate}

All trials were initiated with the subjects holding a static \textit{N-Pose} for 5 seconds and then proceeding to execute the motions. This was always initiated on the same location and as far away as possible to sources of magnetic disturbance for each of the datasets.

Additionally, in D2, a few set of additional trials were collected (1 repetition for each motion, with 5 subjects) with the same protocol, indoors and outdoors, to compare the results obtained in the presence of magnetic disturbances.

%%%%%%%%%%%%%%%%%%%%%%%%%%%%%%%%%%%%%%%%%%%%%%%%%%%%%%%%%%%%%%%%%%%%%%%%%%%%%%%%
%%%%%%%%%%%%%%%%%%%%%%%%%%%%%%%%%%%%%%%%%%%%%%%%%%%%%%%%%%%%%%%%%%%%%%%%%%%%%%%%
\subsection{Dataset elaboration}
\label{sec:methods_dataset}

\subsubsection{Raw data}
\label{sec:methods_dataset_raw}

This protocol was used to collect 2.5M and 1.0M samples of data, respectively in the D1 and D2 datasets, at 60Hz, across all movement sequences (Ergowear data is sampled at around 100Hz, however, the MTwAwinda and \gls{GT} Xsens data are only sampled at 60Hz). Each timestep contains: raw accelerometer, gyroscope and magnetometer data for each of the sensors; along with synchronized data exported from the Xsens MVN Analyze software, including sensor readings (magnetometer, gyroscope, accelerometer, and orientation) in segment referential, center-of-mass location, joint kinematics (angle, velocity, acceleration) and segments' position and orientation (these can be consulted on \cite{xsens_mvn_usermanual}).

%%%%%%%%%%%%%%%%%%%%%%%%%%%%%%%%%%%%%%%%%%%%%%%%%%%%%%%%%%%%%%%%%%%%%%%%%%%%%%%%

% moved here so it appears on desired place on paper
\begin{figure*}[t]
    \centering
    \includegraphics[width=0.95\textwidth]{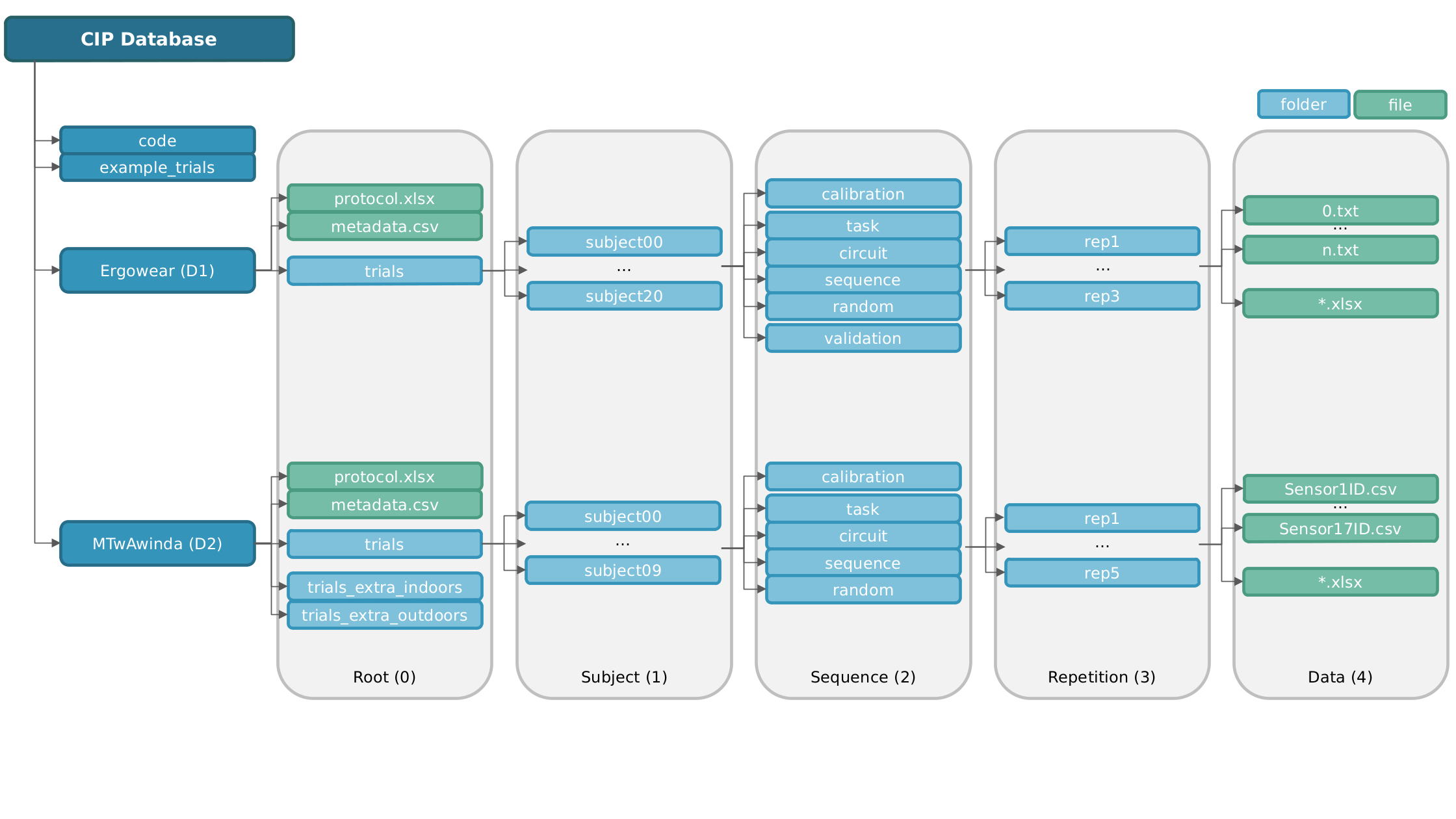}
    \vspace*{-1.4cm}
    \caption{Hierarchical folder structure of the database.}
    \label{fig:database_structure}
\end{figure*}

\subsubsection{Processed data}
\label{sec:methods_dataset_processesing}

In order to validate the acquired data, multiple processing steps were performed, in order to obtain the segments' orientations from the raw sensors' data.

Firstly, all sensors were calibrated to remove bias and scale factors over all axis, following standard methods on \gls{MARG} sensor literature.% \cite{[?]}.

Sensor-to-segment calibration was then performed, transforming the inertial data from the sensors' referential to the body segments' referential. The segment's referential aligns with the world referential in the \gls{NWU} frame of reference when the user holds the T-Pose. This is consistent with the frame of reference used by the \gls{GT} Xsens Analyze software \cite{xsens_mvn_usermanual}, enabling direct comparison of the data. A simple static-calibration \cite{Choe2019s2s} method was used to obtain a reasonable transformation, using the static N-Pose data collected on the start of each trial (first 5 seconds). This method assumes a default sensor placement which is then corrected in the pitch/roll axis using earth's gravity vector as reference.

Finally, the sensor orientations' (now in the segment's referential) were calculated using a standard Madgwick \cite{Madgwick2011} complementary fusion filter.

This data was then used, in conjunction with a kinematic skeleton of the subject, to also obtain keypoint positions through \glspl{FK}, for easy visualization. These data can also be compared to the \gls{GT} data by resampling the data to the same frequency and mapping the corresponding reference segments to match the predicted skeleton.

%%%%%%%%%%%%%%%%%%%%%%%%%%%%%%%%%%%%%%%%%%%%%%%%%%%%%%%%%%%%%%%%%%%%%%%%%%%%%%%%

\subsubsection{Calibration data}
\label{sec:methods_dataset_calibration}

In D2 datasets, offline sensor-to-segment calibration for each subject was performed using an optimization process \cite{nomura2020warmcmaes} to minimize the \gls{MSE} distance between sensor inertial data (accelerometer and gyroscope) and the respective \gls{GT} trajectories, yielding relatively low orientation offsets. This was performed using multiple longer trials, displaying slow dynamics with high variability (sequence and task trials), and resulted in lower error than the standard calibration trials. The transformation was assumed to be constant across trials for the same subject, since the sensors were firmly secured with straps with good grip.

Unfortunately, this approach did not work reliably for the Ergowear data, since there was some sensor unpredictable displacement across trials. This derived from the subject's movement and the fact that the smart-garment prototype was not capable to firmly secure the sensors to the subjects' segments.

%%%%%%%%%%%%%%%%%%%%%%%%%%%%%%%%%%%%%%%%%%%%%%%%%%%%%%%%%%%%%%%%%%%%%%%%%%%%%%%%%%%%
%%%%%%%%%%%%%%%%%%%%%%%%%%%%%%%%%%%%%%%%%%%%%%%%%%%%%%%%%%%%%%%%%%%%%%%%%%%%%%%%%%%%
%%%%%%%%%%%%%%%%%%%%%%%%%%%%%%%%%%%%%%%%%%%%%%%%%%%%%%%%%%%%%%%%%%%%%%%%%%%%%%%%%%%%

\section{Data Records}
\label{sec:records}

All the collected data was organized in a database to enable reuse across the research community \cite{palermo2022cipdatabase}, and contains both \emph{Ergowear} and \emph{MTwAwinda} trials, i.e., D1 and D2 datasets, along with code to handle the data.

Each of the datasets are structured hierarchically in 5 levels (Figure \ref{fig:database_structure}), following a similar organization to provide an intuitive and easy way to select desired data:
\textbf{i)} level 0: \textbf{Root}, participant's metadata, acquisition protocol, general dataset information and a folder containing the trials.
\textbf{ii)} level 1: \textbf{Subject}, a folder for each of the subjects and calibration files;
\textbf{iii)} level 2: \textbf{Sequence}, contains a folder for each type of sequences performed (calibration, task, circuit, sequence, random and validation);
\textbf{iv)} level 3: \textbf{Repetition}, includes a folder with the repetitions id and possible annotations (ex. identifying data corruption or longer trials);
and \textbf{v)} level 4: \textbf{Data}, presents the files containing the trial data and synchronized \gls{GT}.

%%%%%%%%%%%%%%%%%%%%%%%%%%%%%%%%%%%%%%%%%%%%%%%%%%%%%%%%%%%%%%%%%%%%%%%%

\subsection{D1 - Ergowear data}
\label{sec:records_dataset_ergowear}
Each of the D1 trial folders contain two types of data files, the first acquired with \textit{ad-hoc} Ergowear software and the second, the Xsens data exported using the proprietary Xsens MVN Analyze software.

The Ergowear files are stored as plain text (\textit{(.txt)} files, with multiple columns: The packet index, \gls{MARG} (accelerometer, gyroscope and magnetometer) data over the 3 axis, for each of the 9 \glspl{IMU} and a relative timestamp of the packets. Each file contains only around 3000 packets, with more samples being divided into multiple files. These are named in integer order (i.e. 1.txt, 2.txt, 3.txt, ...). The data is sampled at 100 Hz.

The Xsens data is contained in an excel file (\textit{.xlsx}), exported from the Xsens Analyze software, which can be consulted on the user manual \cite{xsens_mvn_usermanual}. It contains multiple sheets each with a different data modality, which not only includes the inertial sensor information, but also segments’ orientation, position, velocity, and acceleration; and joints’ angle. This data is sampled at 60Hz.

\subsection{D2 - MTwAwinda data}
\label{sec:records_dataset_mtwawinda}
Each of the D2 trial folder also contains the Xsens Analyze files as above, and data from the second set of Xsens MTw Awinda sensors, using the Xsens MtManager software for the data acquisition. To avoid misunderstanding of the terms, this paper uses MTwAwinda data as reference for the raw data acquired with Xsens MtManager software and Xsens Analyze data as \gls{GT} data.

The MTwAwinda files are stored as (\textit{.csv} files, one for each of the sensors (17 sensors), as exported by the MtManager software, and each named with the respective sensor ID. Each file contains columns with the following data: The packet index, \gls{MARG} (accelerometer, gyroscope and magnetometer) data over the 3 axis and the sensor orientation quaternion (in sensor referential).

%%%%%%%%%%%%%%%%%%%%%%%%%%%%%%%%%%%%%%%%%%%%%%%%%%%%%%%%%%%%%%%%%%%%%%%%%%%%%%%%%%%%
%%%%%%%%%%%%%%%%%%%%%%%%%%%%%%%%%%%%%%%%%%%%%%%%%%%%%%%%%%%%%%%%%%%%%%%%%%%%%%%%%%%%
%%%%%%%%%%%%%%%%%%%%%%%%%%%%%%%%%%%%%%%%%%%%%%%%%%%%%%%%%%%%%%%%%%%%%%%%%%%%%%%%%%%%
\section{Technical Validation}
\label{sec:validation}

The subjects were instructed to follow the protocol proposed, and incorrect trials were repeated (e.g. performing the sequence trial incorrectly). At the same time, they were instructed to perform the motions as freely and naturally as possible to avoid biasing he results. The responsible researcher supervised and guided the experiments.

Data from both datasets was visually and qualitatively inspected, both online during the acquisition and offline, in order to ensure its validity. Trials containing issues (e.g. data desynchronization, sensors dropping, etc...) were either repeated or discarded.

%%%%%%%%%%%%%%%%%%%%%%%%%%%%%%%%%%%%%%%%%%%%%%%%%%%%%%%%%%%%%%%%%%%%%%%%%%%%%%%%%%%%%%%%%%%%
\subsection{Baseline Results}
\label{sec:validation_results}

\begin{figure}[t]
    \centering
    \begin{subfigure}[b]{0.48\textwidth}
        \centering
        \includegraphics[width=6.5cm]{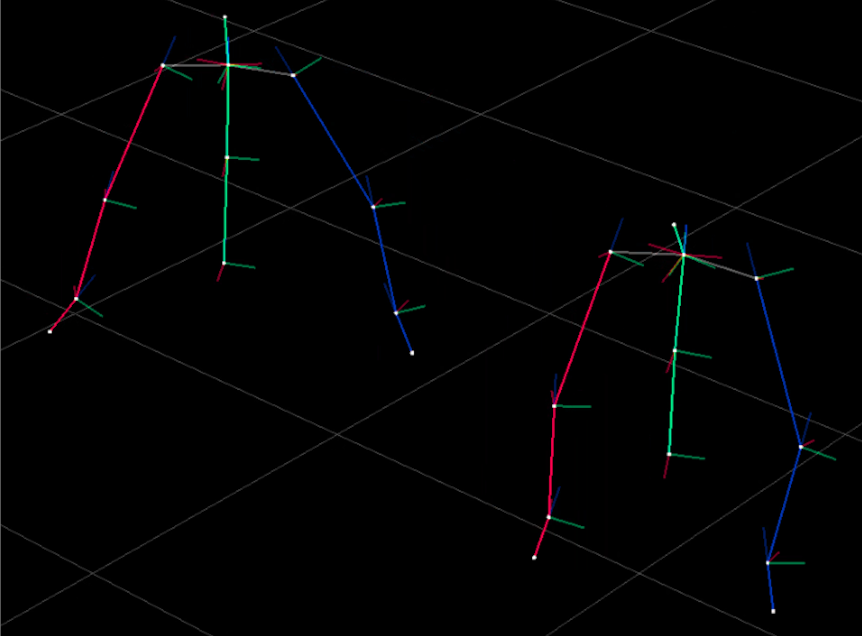}
        %\caption{Ergowear}
    \end{subfigure}
    \begin{subfigure}[b]{0.48\textwidth}
        \centering
        \includegraphics[width=6.5cm]{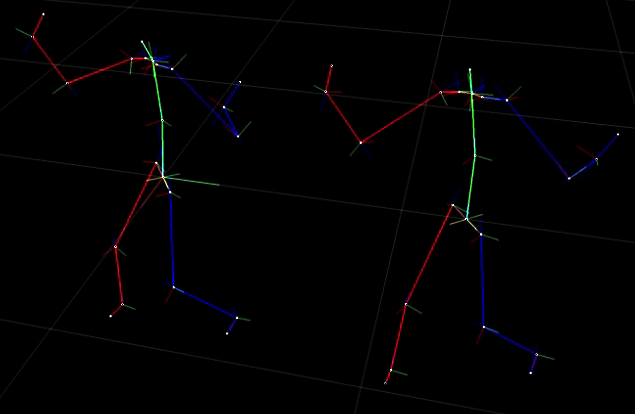}
        %\caption{MTwAwinda}
    \end{subfigure}
    \caption{3D skeleton visualization of the acquired data in the Ergowear (top) and MTwAwinda (bottom) datasets. Each one is compared to the Xsens \gls{GT} (on the left in both cases), which is mapped to each skeleton.}
    \label{fig:results_skeletons}
\end{figure}

Following the method described in Section \ref{sec:methods_dataset_processesing}, the resulting skeletons can be visualized in 3D in Figure \ref{fig:results_skeletons}, where the acquired data displays similar motion as the \gls{GT}.

The average error obtained through the simple method used is presented on Table \ref{tab:table_filter_persequence_results}, for each type of sequence and for each dataset. An overall average error of 27.46º and 7.83º were respectively obtained for the Ergowear and MTwAwinda datasets.

\begin{table}[t]
    \centering
    \caption{Madgwick fusion error compared to \gls{GT} results for each of the sequences on Ergoaware and MTwAwinda data. The Quaternion Angle Distance (QAD) \cite{huynh2009qad} [0, 180º] was used as metric.}
    \maxsizebox{\textwidth}{!}{
    \begin{tabular}{l | c | c }
        Filter          &  Ergowear QADº        &  MTwAwinda QADº  \\ \hline \hline
        Task            &  25.79                &  5.26            \\            
        Circuit         &  26.60                &  7.98            \\            
        Sequence        &  26.77                &  6.36            \\            
        Random          &  30.42                &  9.67            \\ \hline
        Average         &  27.46                &  7.83            \\            
        \hline
    \end{tabular}
    }
    \label{tab:table_filter_persequence_results}
\end{table}

%%%%%%%%%%%%%%%%%%%%%%%%%%%%%%%%%%%%%%%%%%%%%%%%%%%%%%%%%%%%%%%%%%%%%%%%%%%%%%%%%%%%%%%%%%%%
\subsection{Data Limitations}
\label{sec:validation_limitations}

% \begin{itemize}
    % \item Lost data packages.
    % \item Corrupted sensor samples.
    % \item Magnetic interference.
    % \item Sensor displacement artifacts.
    % \item Inconsistent sampling frequency (95-105Hz).
    % \item Relatively high gyroscope bias, leading to high drift when not doing data fusion with magnetometer.
    % \item High offset error from the sensor to segment calibration methods used.
% \end{itemize}

Data inspection reveals high correlation between the xsens \gls{GT} and the collected data (Baseline Results Section). However, some issues were noticed, which will inevitably lead to angle estimation errors. These were more pronounced in the D1, with Ergowear acquisition, and minimized as much as possible in the D2, with MTwAwinda acquisition.
\textbf{i)} trials contain lost data packets or occasional corrupted samples, which can vary from 0\% to as high as 9.5\% in a few trials;
\textbf{ii)} presence of magnetic disturbances thorough the trials, both from the surrounding environment (e.g. cabinets, table legs, building structure and wiring, etc... in the case of indoors acquisition)) and interacted objects (e.g. chair, laptop, stool, in both acquisitions);
\textbf{iii)} both datasets contain entire trials which have been corrupted (and thus discarded), this amounts to 35/1197 in the case of D1 and 43/340 in the case of D2;
\textbf{iv)} presence of movement and soft-body artifacts, which although reduced as much as possible (especially in the D2 dataset), cannot be completely removed;
\textbf{v)} the commercial \gls{MoCap} system used as \gls{GT}, although validated, is still dependent on good calibration and is affected by magnetic interference, and will thus contain higher errors than traditional visual \gls{MoCap} systems \cite{AlAmri2018xsenserror}, adding to computed errors.

Additionally, the D1, concerning Ergowear acquisition, given its prototype state, presents additional issues:
\textbf{i)} inconsistent sampling frequency, which might oscillate around the expected 100Hz ($\sim$95-105Hz);
\textbf{ii)} relatively high gyroscope bias instability, caused by the fact that low-cost \gls{MARG} sensors were used (MPU9250) and absence of Strap-Down Integration, leading to lower quality gyroscope estimates;
\textbf{iii)} sensor displacement artifacts, caused by the fact that the Ergoaware sensors were placed over the jacket, which although tight, might slide, especially during near maximal extension, leading to variable sensor-to-segment offsets across timesteps and different trials;
\textbf{iv)} there are no shoulder sensors, which might be necessary for a better assessment of some upper-body movements.

%%%%%%%%%%%%%%%%%%%%%%%%%%%%%%%%%%%%%%%%%%%%%%%%%%%%%%%%%%%%%%%%%%%%%%%%%%%%%%%%%%%%
%%%%%%%%%%%%%%%%%%%%%%%%%%%%%%%%%%%%%%%%%%%%%%%%%%%%%%%%%%%%%%%%%%%%%%%%%%%%%%%%%%%%
%%%%%%%%%%%%%%%%%%%%%%%%%%%%%%%%%%%%%%%%%%%%%%%%%%%%%%%%%%%%%%%%%%%%%%%%%%%%%%%%%%%%

\section{Code Availability}
\label{sec:code}

% \begin{itemize}
%     \item Statement on how the code to use the data can be used
% \end{itemize}

This database is accompanied by a folder with all the scripts used to process and handle the data described. It is openly hosted in \emph{Zenodo} \cite{palermo2022cipdatabase}.

Additionally, an extended code repository is available on Github\footnote{\url{https://github.com/ManuelPalermo/HumanInertialPose.git}} with updated code to not only process the data described, but also calculate kinematics, visualize and evaluate the resulting motions and offers extended support for general inertial pose estimation pipelines. All scripts are based on the Python programming language and, thus, open source. The code contains a permissive MIT license for unrestricted usage.

We hope this dataset and associated code can further contribute to the development and evaluation of classic or data-driven inertial human pose estimation solutions, with applications, for example, in human movement understanding and forecasting, ergonomic assessment and gait/posture analysis.

%%%%%%%%%%%%%%%%%%%%%%%%%%%%%%%%%%%%%%%%%%%%%%%%%%%%%%%%%%%%%%%%%%%%%%%%%%%%%%%%%%%%
%%%%%%%%%%%%%%%%%%%%%%%%%%%%%%%%%%%%%%%%%%%%%%%%%%%%%%%%%%%%%%%%%%%%%%%%%%%%%%%%%%%%

\section*{Acknowledgements}
\label{sec:acknowledgements}

\noindent This work is supported by: European Structural and Investment Funds in the FEDER component, through the Operational Competitiveness and Internationalization Programme (COMPETE 2020) [Project nº 39479; Funding Reference: POCI-01-0247-FEDER-39479]. Sara Cerqueira was supported by the doctoral Grant SFRH/BD/151382/2021, financed by the Portuguese Foundation for Science and Technology (FCT), under MIT Portugal Program.

\section*{Author contributions statement}
\noindent M.P., S.C., J.A., and C.P.S. conceived the data acquisition; M.P., S.C., J.A. and A.P. conducted the data acquisition; M.P. analyzed and processed the data; M.P. organized the database; M.P., S.C. and C.P.S. contributed to the manuscript's edition; All authors reviewed the manuscript.

%%%%%%%%%%%%%%%%%%%%%%%%%%%%%%%%%%%%%%%%%%%%%%%%%%%%%%%%%%%%%%%%%%%%%%%%%%%%%%%%%%%%
%%%%%%%%%%%%%%%%%%%%%%%%%%%%%%%%%%%%%%%%%%%%%%%%%%%%%%%%%%%%%%%%%%%%%%%%%%%%%%%%%%%%
%%%%%%%%%%%%%%%%%%%%%%%%%%%%%%%%%%%%%%%%%%%%%%%%%%%%%%%%%%%%%%%%%%%%%
\ifCLASSOPTIONcaptionsoff
  \newpage
\fi

\printbibliography[heading=bibintoc, title={References}]

\end{document}